# Superconductivity at 33 K in potassium-doped 1,2:8,9-dibenzopentacene

Mianqi Xue[†,‡], Tingbing Cao[‡], Duming Wang[§], Yue Wu[†], Huaixin Yang[†], Xiaoli Dong[†], Junbao He[§], Fengwang Li[‡], G. F. Chen[†,§]

[†]Institute of Physics and Beijing National Laboratory for Condensed Matter Physics, Chinese Academy of Sciences, Beijing 100190, China

[‡]Department of Chemistry, Renmin University of China, Beijing 100872, China

[§]Department of Physics, Renmin University of China, Beijing 100872, China

Recently, superconductivity at 18 K was reported in potassium-doped picene, a hydrocarbon molecular whose five benzene rings are condensed in an armchair manner[1]. $K_3$phenanthrene, composed of three fused benzene rings, was subsequently synthesized and found to be superconducting at 5 K [ref. 2]. These observations suggested that the superconducting transition temperature is clearly dependent on the length of polycyclic-aromatic-hydrocarbon (PAH) chains. It remains to be seen if a much higher transition temperature $T_c$ than 18 K can be achieved in PAHs with different number of benzene rings. Here we report the observation of superconductivity at 33 K in K-doped 1,2:8,9-dibenzopentacene ($C_{30}H_{18}$). This is higher than any $T_c$ reported previously for an organic superconductor besides the alkali-metal doped $C_{60}$ [ref. 3]. This finding provides an indication that superconductivity at much higher temperature may be possible in such system, if PAHs with longer benzene rings can be synthesized.

Samples were prepared by direct heating potassium metal with 1,2:8,9-dibenzopentacene in an evacuated tube at 300-350 ℃ for 7-20 days. To improve the homogeneity of products, a second anneal is sometimes performed. An alternative method for sample synthesis was also developed. The starting materials were loaded into liquid ammonia solution and stirred for 6 hours. The resulting products were then dried under vacuum for several hours to remove the solvent, and subsequently sealed in an evacuated tube and annealed at 250-300 ℃ for 7-20 days. The latter synthetic route allows for a reduction of annealing temperature and improvement of homogeneity of the products, but it does not allow precise control of the potassium content of the resulting samples. All the obtained products are uniform dark black in color, which is totally different from the red color of pure dibenzopentacene. Superconductivity was observed for the samples with compositions of the form $K_x$dibenzopentacene, for $3.0 \leqslant x \leqslant 3.5$. It seems to be the fact that the superconducting shielding fraction could be improved by increasing the annealing temperature.

The temperature dependence of the magnetization measured in a field of 10 Oe in a SQUID magnetometer (Quantum Design) for a sample of composition $K_{3.17}$dibenzopentacene is given in Fig. 1. The zero-field-cooling (ZFC) and field-cooling (FC) susceptibility shows a sharp drop at around 28.2 K which also can be seen from the ac susceptibility as shown in the left inset in Fig. 1a. Such diamagnetic behavior is characteristic of superconductivity. The distinct magnetic signatures of ZFC and FC below the superconducting critical temperature originate from the screening supercurrents (ZFC regime) and the Meissner-Ochsenfeld effect of magnetic flux expulsion (FC regime). The diamagnetic signal onset temperature is described as superconductivity temperature $T_c^{onset}$. As seen from Fig. 1a, $T_c^{onset}$ was defined to be 28.2 K for $K_{3.17}$dibenzopentacene. The shielding volume fraction at 5 K is estimated to be 5.5% (assuming a density of 1.8 g/cm$^3$), which is comparable to the value reported for superconducting $K_x$picene[1], $K_x$phenanthrene[2], and is about the same as that initially reported for K-doped $C_{60}$ and for Rb-doped $C_{60}$ [ref. 4, 5], although following works have dramatically increased the superconducting fraction for alkaline-earth-metal doped phenanthrene[6] and alkali-metal doped

$C_{60}$ [ref. 7, 8]. The small superconducting fraction may be due to the smaller sizes of the obtained crystallites than the London penetration depths[1]. Further investigations of the dependence of the superconducting fraction on temperature, initial composition and reaction time will undoubtedly lead to high yields of superconducting phase.

Figure 1b shows temperature dependence of measured $\chi$ for various applied fields $H$ in the ZFC measurements. There was an obvious drop of $\chi$ at 23 K even at 1000 Oe, indicating that the superconducting phase is not completely destroyed at weak applied field. The fact that $T_c$ is suppressed slowly by applying the magnetic fields indicates the observed superconductivity in this material is intrinsic. The corresponding upper critical field $H$ versus $T_c$ is plotted in Fig. 1c. $T_c$ was determined through a linear extrapolation of the slopes before and after the point at which the sample began superconducting, as shown in Fig. 1b. At the present stage, it is difficult to determine the upper critical field $H_{c2}$ from $H$-$T_c$ curve. Figure 1d depicts the $M(H)$ versus $H$ plot at 4.5 K measured by sweeping the magnetic field at a constant rate of 10 Oe/sec, which indicates that $K_{3.17}$dibenzopentacene is a type-II superconductor with a strong vortex pinning. The lower critical field $H_{c1}$ was estimated to be 200 Oe at 4.5 K. which is higher than that of the potassium doped picene with $T_c$ = 18 K [ref. 1] ($H_{c1}$ ~100 Oe; however, the authors gave the value of 380 Oe determined from the minimum position of the $M$-$H$ plot).

Superconductivity with higher temperature can also be observed for high K-content sample, $K_{3.45}$dibenzopentacene. The onset temperature of diamagnetism is $T_c^{onset}$ = 33.1 K as seen in ZFC experiment (see Fig. 2a), where the critical temperature is the highest to date among the organic superconductors besides the alkali-metal doped $C_{60}$ [ref. 3]. The magnitude of the shielding signal at 25 K corresponds roughly to 3.2% of perfect diamagnetism. Additional features are seen at 20 K and 5 K, which might be ascribed to different superconducting phases with lower transition temperatures. One should note that the latter two characteristic temperatures are similar to the reported values for the superconducting $K_x$picene and $K_x$phenanthrene[1,2]. It is reasonable to suspect that these two superconducting phases come from the breakdown of PAH chains during the reaction process. The $T_c$ obtained for twice annealed sample of $K_3$dibenzopentacene was 7.4 K, with superconducting shielding fraction of 3.6%, as shown in Fig. 2b. Similar phenomena has been observed in $K_x$picene[1], where two superconducting phases ($T_c$ = 7 or 18 K) were occasionally obtained under the same experimental conditions, but did not coexist with each other.

For non-superconducting $K_x$dibenzopentacene, we find that the magnetic susceptibility shows a Pauli- or Curie-like behavior, indicating that there exists local spins in this material. Similar behaviors have also been observed in $K_x$picene and $K_x$phenanthrene[1,2]. The origin of spin is still not clear. Recent density functional calculations suggested that $K_x$piecene superconductor was near the metal-insulator transition (MIT) and the magnetic instability[9]. Therefore, the Pauli- or Curie-like magnetization behaviors observed in non-superconducting $K_x$dibenzopentacene might suggest a role for spin fluctuations in the pairing mechanism in this material. In fact, superconductivity competing with spin density wave (SDW) antiferromagnetism has been reported in lots of organic superconductors[10,11], and high $T_c$ cuprates[12,13] and recently discovered iron pnictides[14,15].

1,2:8,9-dibenzopentacene is a condensed aromatic hydrocarbon with both linear and angular fusion of the benzene rings, which can be viewed as two phenanthrene segments are bridged through two –CH= group in an anti arrangement (see right inset of Fig. 1a). Regardless of the influence of the structural/geometric difference among picene, phenanthrene and dibenzopentacene (one should note that superconductivity with $T_c$ up to 15 K was also observed for K-doped coronene, which is composed of six peri-fused benzene rings. It is still not clear whether the perfect W-shaped configurations of benzene rings are the key role to achieve superconductivity in doped PAHs[9,16-18], although there is no superconductivity found for petancene, the isomer of picene[1]), it can be seen that with increasing the length of PAH chain, the superconducting transition temperature increases dramatically (see Fig. 3): $T_c$ increases from 5 K for $K_x$phenanthrene[2] with three benzene rings to 18 K for $K_x$picene[1] with five benzene rings, and up to 33.1 K for $K_x$dibenzopentacene with seven benzene rings, which is in contrast to the theoretical predication[19,20]. In general, the dimensionless electron-phonon coupling constant $\lambda_x$ [$=N(\varepsilon_F)l_x$] is directly related to $l_x$ and $N(\varepsilon_F)$. Here the total electron-phonon coupling constant $l_x$ for the molecule, dominated by intramolecular vibrational modes, is inversely proportional to the number of atoms in the molecule that donates π-electrons. Assuming a conventional Bardeen-Cooper-Schrieffer (BCS) mechanism with electron–phonon interaction, and the same density of states $N(\varepsilon_F)$ at the Fermi level, the doped phenanthrene should have the highest $T_c$ in all the aromatic hydrocarbons[19,20]. However, it should be noted that, in a molecular crystal, there are several factors that influence the properties, including the length of the chain (here the number of benzene rings), the three-dimensional arrangement of the chains, the strength/types of intermolecular forces between the chains and the orientation of benzene rings within the chain. Usually, increasing the length of the PAH chain would increase the extent of its interactions with neighboring chains, while the density of states at the Fermi level, is mainly dominated by intermolecular interactions. Therefore, the superconducting temperature $T_c$ depends on the relative magnitude of the strength of $l_x$ and $N(\varepsilon_F)$. One can speculate that an increase in $N(\varepsilon_F)$ overcomes any decreases in $l_x$ on going from $K_3$phenanthrene to $K_3$dibenzopentacene.

In summary, we have observed superconductivity at temperature as high as 33.1 K in a new K-doped hydrocarbon, 1,2:8,9-dibenzopentacene. Our result, combined with previous reports, indicates that the superconducting transition temperature is clearly dependent on the length of PAH chain. The increase of the chain length will lead to stronger intermolecular interaction in the solid state. One might expect, high temperature superconductivity with a transition temperature significantly exceeding that of intercalated $C_{60}$, could be realized when the length of PAH chain is long enough. We suggest the PAHs represent one road to high $T_c$ that is worthy of further exploration.

**Acknowledgements**

The authors wish to thank Z. X. Zhao, L. L. Sun for their fruitful discussions and helpful comments. This work was supported by the Natural Science Foundation of China, and by the Ministry of Science and Technology of China.

**Figure Legends**

**Figure 1: Temperature dependence of magnetization for $K_{3.17}$dibenzopentacene. a**, $\chi$ versus $T$ plots for $K_{3.17}$dibenzopentacene with $T_c^{onset}$ = 28.2 K at $H$ = 10 Oe (main panel). The left inset shows the ac susceptibility at $H$ = 5 Oe, and the right inset shows the molecular structure of dibenzopentacene; unfortunately, to our knowledge, no report on the crystal structure of 1,2:8,9-dibenzopentacene has been published. **b**, $\chi$ versus $T$ plots for the sample in the ZFC measurements under different magnetic field $H$. The solid lines indicate the determination of the transition temperature as described in the text. **c**, The $H$ versus $T_c$ plot. **d**, $M$ versus $H$ plots for the sample at 4.5 K. The solid line is a guide for the eye.

**Figure 2: Temperature dependence of magnetization for $K_{3.45}$dibenzopentacene and $K_3$dibenzopentacene. a**, $\chi$ versus $T$ plots for superconducting $K_{3.45}$dibenzopentacene at $H$ = 10 Oe (main panel). The ZFC curve indicates the presence of three different superconducting phases. The left inset shows on an expanded scale, the region around $T_c$, revealing a critical temperature of 33.1 K. The right inset shows $\chi$ versus $T$ plots for the sample $K_{3.45}$dibenzopentacene in the ZFC measurements under different magnetic field $H$. It clearly shows that all the three superconducting transitions are suppressed slowly by applying the magnetic fields. **b**, $\chi$ versus $T$ plots for the twice annealed $K_3$dibenzopentacene sample with $T_C$ = 7.4 K at $H$ = 10 Oe.

**Figure 3: Plots of the superconducting transition temperature versus the number of benzene rings.** The transition temperature increases from 5 K to 18 K, and then up to 33 K which is linear relatively to the number of benzene rings.

**Figure 1**

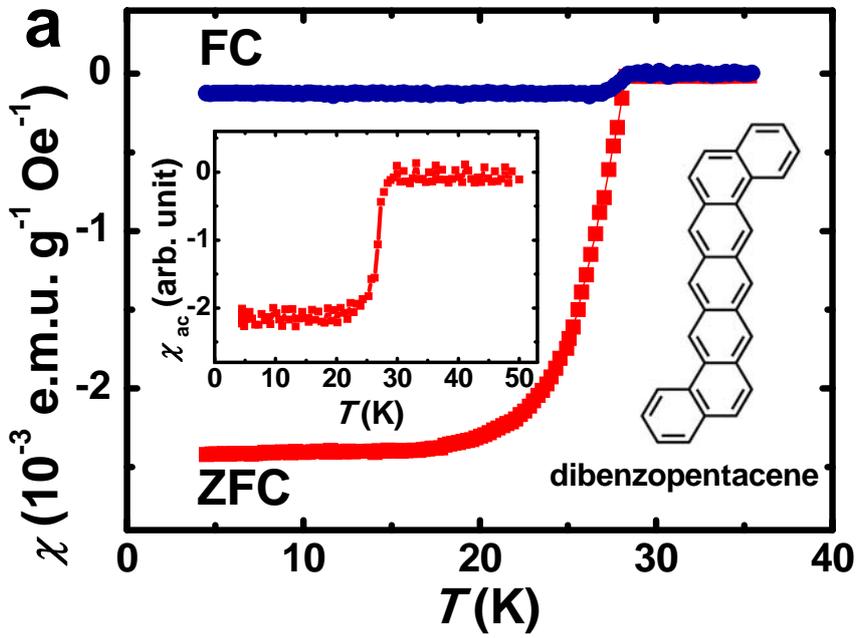

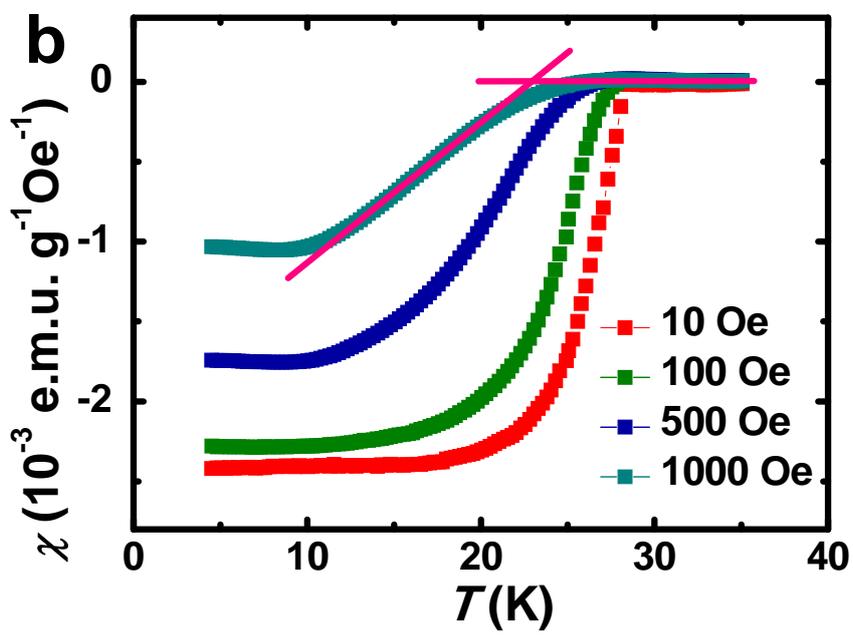

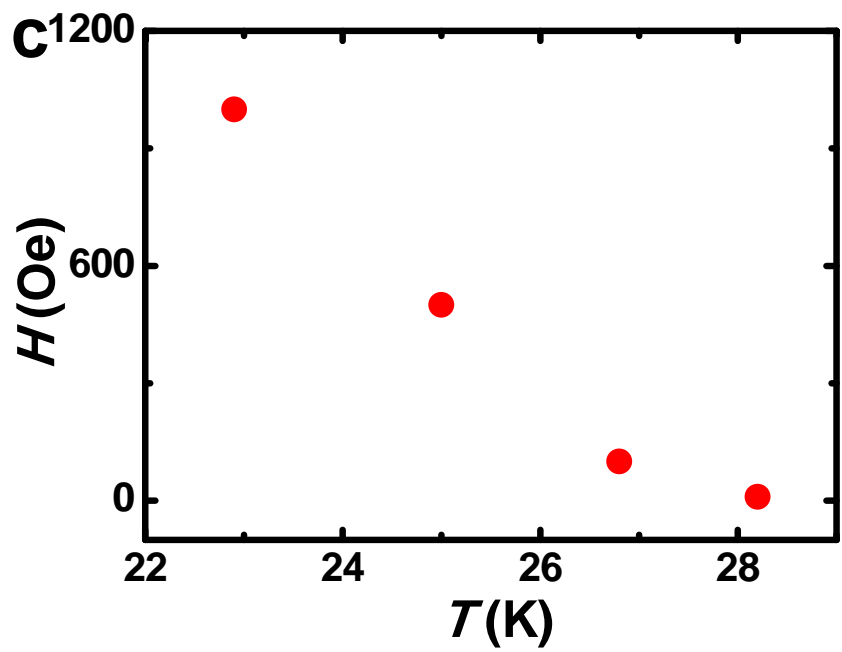

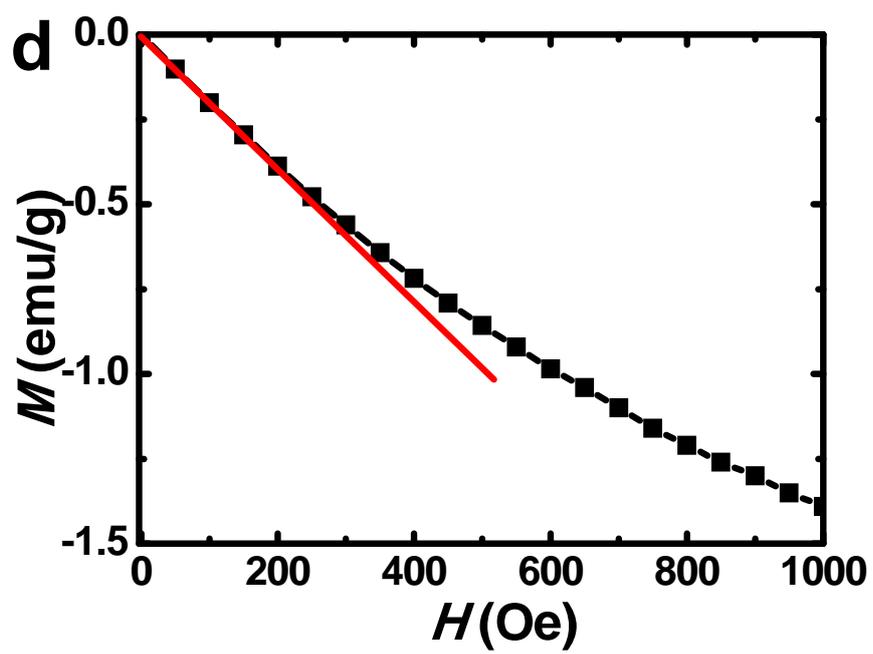

**Figure 2**

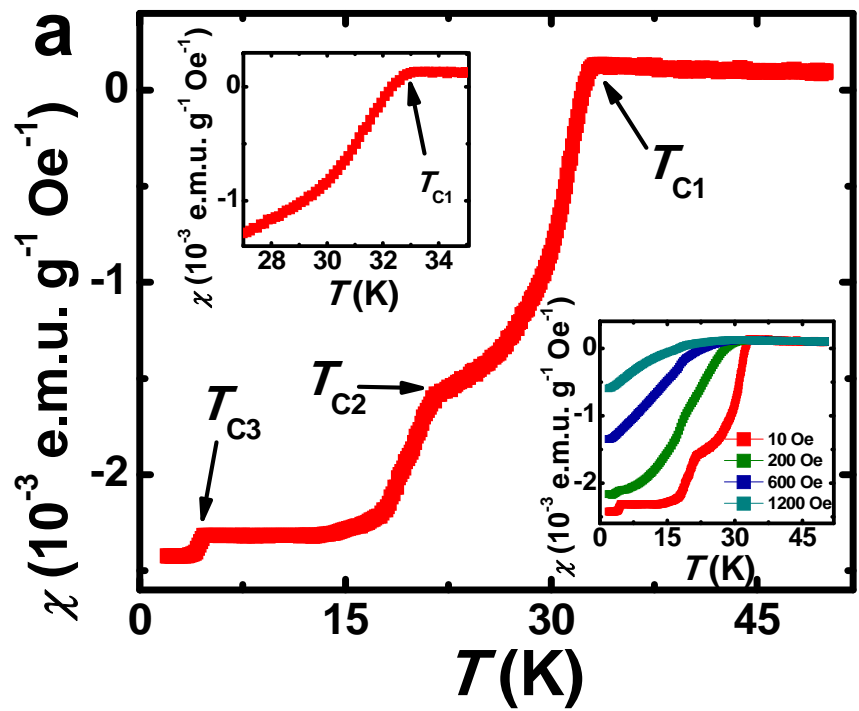

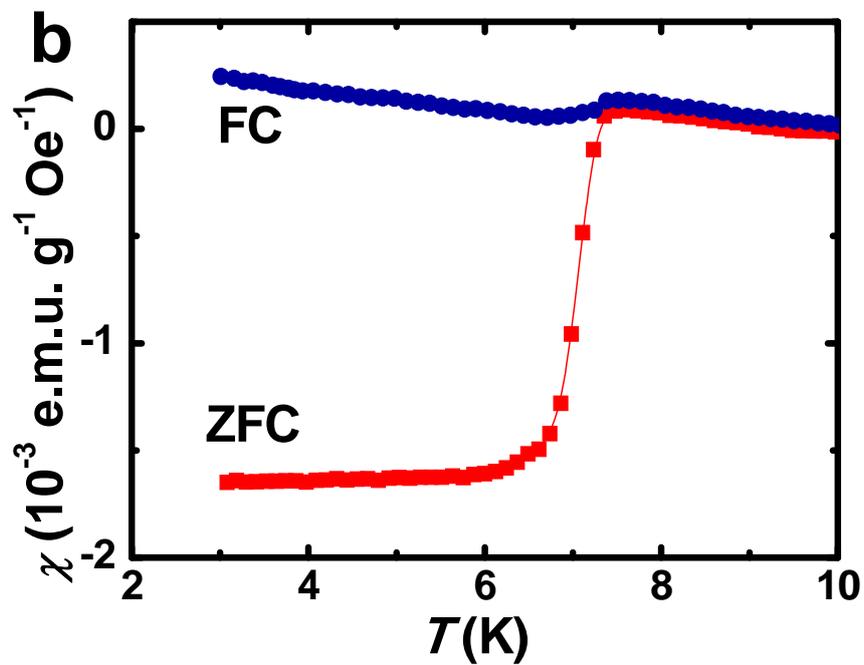

Figure 3

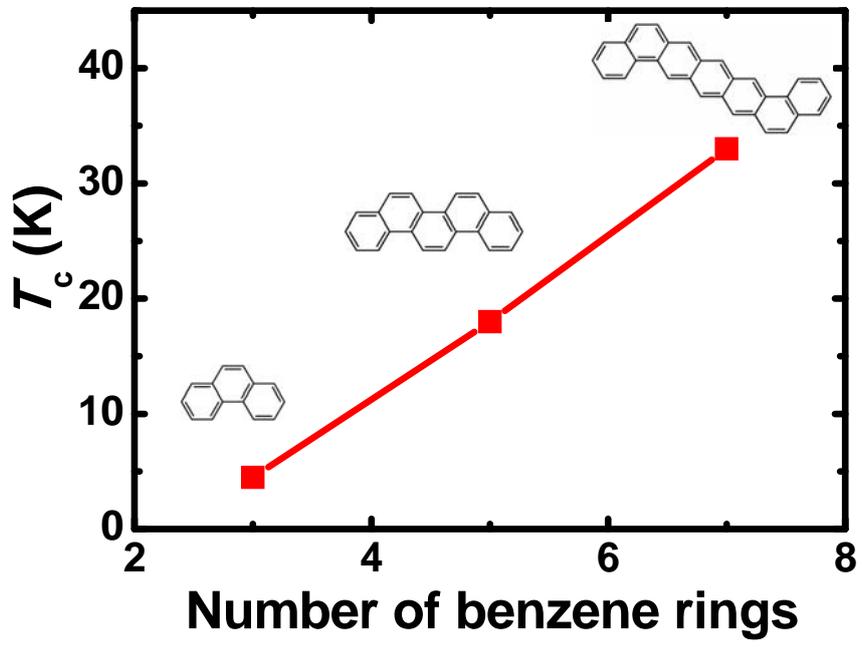